\newcommand{\DVSRWELL}{$\rm{\Delta V_{SRWELL}}$}
\newcommand{\DVDRIFT}{$\rm{\Delta V_{Drift}}$}
\newcommand{\DVTRANSFER}{$\rm{\Delta V_{Transfer}}$}
\newcommand{\DVTHGEM}{$\rm{\Delta V_{THGEM}}$}
\newcommand{\kHzcm}{$\rm{kHz \over cm^2}$}
\newcommand{\Hzcm}{$\rm{Hz \over cm^2}$}

\documentclass{JINST}

\title{Beam studies of novel THGEM-based potential sampling elements for Digital Hadron Calorimetry }

\author{S. Bressler$^a$, L. Arazi$^a$, H. Natal da Luz$^b$, C. D. R. Azevedo$^c$, L. Moleri$^a$, E. Oliveri$^d$, M. Pitt$^a$, A. Rubin$^a$, J. M. F. dos Santos$^b$, J. F. C. A. Veloso$^c$ and A. Breskin$^a$\\
\llap{$^a$}Dept. of Astrophysics and Particle Physics, Weizmann Institute of Science,\\
  P.O. Box 26 Rehovot 76100, Israel\\
\llap{$^b$}Physics Department, FCTUC, University of Coimbra, \\
3004-516 Coimbra, Portugal \\
\llap{$^c$}I3N, Physics Department, University of Aveiro, \\
3810-193 Aveiro Portugal \\
\llap{$^d$}CERN,\\
  Meyrin, Switzerland\\
  E-mail: \email{shikma.bressler@weizmann.ac.il}}

\abstract{Beam studies of thin single- and double-stage THGEM-based detectors are presented. Several $10 \times 10~\rm{cm^2}$ configurations with a total thickness of 5-6 mm (excluding readout electronics), with $1 \times 1~\rm{cm^2}$ pads inductively coupled through a resistive layer to APV-SRS readout electronics, were investigated with muons and pions. Detection efficiencies in the 98\% range were recorded with an average pad-multiplicity of $\sim$1.1. The resistive anode resulted in efficient discharge damping, with few-volt potential drops; discharge probabilities were $\sim 10^{-7}$ for muons and $10^{-6}$~ for pions in the double-stage configuration, at rates of a few \kHzcm. These results, together with the robustness of THGEM electrodes against spark damage and their suitability for economic production over large areas, make THGEM-based detectors highly competitive compared to the other technologies considered for the SiD-DHCAL.}


\begin{document}

\section{Introduction}

The Thick Gas Electron Multiplier (THGEM)~\cite{Chechik2004} is a simple, robust and economic detector element; it can be industrially produced over large areas using standard Printed Circuit Board (PCB) technologies. Its properties and potential applications are reviewed in~\cite{Breskin2009,Breskin2011}; for recent works on THGEM properties in normal operation conditions see ~\cite{Peskov2010,Alexeev2012,Azavedo2010,Alexeev2013}. In the last few years, a considerable R\&D effort was undertaken to evaluate the applicability of THGEM-based detectors as thin sampling elements in Digital Hadronic Calorimeters (DHCAL). The effort focused, in particular, on the DHCAL of the Silicon Detector (SiD)~\cite{Burrows}. However, since THGEM detectors provide proportional response, they are also applicable in the semi-DHCAL concept~\cite{Brau2013}; in the latter, different threshold levels applied to the deposited-charge signals should permit reducing hadronic-response non-linearities.

SiD is one of the two detector concepts studied for future linear collider, either the International Linear Collider (ILC)~\cite{Brau2013} or the Compact Linear Collider (CLIC)~\cite{Linssen2012}. In its present design, the SiD DHCAL will require a total active area of a few thousand square meters of a few-millimetres thick sampling elements~\cite{Burrows}. The expected average particle rates at the DHCAL are of $\sim$1 \kHzcm. 

The ambitious physics program of both the ILC and CLIC requires excellent, 3-4\%, jet energy resolution (defined as $\rm{\sigma_E \over E}$, where E is the true jet energy and $\rm{\sigma_E}$~is the resolution of the energy measurement). In order to achieve this resolution, the SiD DHCAL is designed to employ Particle Flow Algorithms (PFAs)~\cite{Xia2006}, and is thus highly segmented in both the longitudinal and transverse directions. Its baseline design comprises of 40-50 layers of absorber plates (either stainless steel or tungsten) separated by 8 mm gaps, which should incorporate the active sampling elements with their $1 \times 1~\rm{cm^2}$~ pixels and readout electronics~\cite{Burrows}. High single-particle detection efficiency and low average pad-multiplicity (number of pads activated per crossing particle) are essential in this application.

Active elements utilizing 3-5 mm thick Resistive Plate Chambers (RPCs), excluding readout electronics, are the baseline technology for the SiD hadronic calorimeter~\cite{Drake2007,Bilki2009a}. They were already tested at the level of a $1~\rm{m^3}$~instrumented block with digital readout electronics and yielded so far an average pad multiplicity of $\sim$1.6 at 94\% efficiency with muons~\cite{Bilki2009b}. Detection elements based on the MICRO MEsh GAseous Structure (MICROMEGAS), tested with muons at $1 \rm{m^2}$~units, displayed superior properties (with a thickness of $\sim$3.1 mm, excluding the MICROROC electronics~\cite{Adloff2012}): 97\% efficiency with a 1.1 average multiplicity~\cite{Adloff2009}. Detector prototype using $30 \times 30~\rm{cm^2}$~double Gas Electron Multipliers (double-GEMs) of 5 mm thickness excluding the KPiX readout electronics~\cite{Freytag2008}, yielded a multiplicity of ~1.3 at 95\% efficiency with muons~\cite{Yu2012}.
 
THGEM-based structures were proposed as potential DHCAL sampling elements in 2010 and have since been thoroughly investigated in laboratory and beam settings. The results of previous beam studies, performed at the RD51 test-beam facility of CERN-SPS, are summarized in~\cite{Arazi2012}.  They focused primarily on single- and double-THGEM detector structures of $10 \times 10~\rm{cm^2}$, operated in Ne/5\%CH4; the avalanche charge was recorded on $1 \times 1~\rm{cm^2}$~anode readout pads following a 2 mm induction gap and directly coupled to KPiX readout electronics ~\cite{Freytag2008}. These preliminary studies demonstrated the potential value of such detectors for DHCAL, with an average pad multiplicity of $\sim$1.1-1.2 at a muon detection efficiency of $\sim$95\% for a $\sim$6 mm thick single-stage detector, in which charges were deposited in a 4 mm thick drift gap. These encouraging results left, however, room for optimization, in terms of detector thickness and stability in harsh hadronic environment. 

\begin{figure}
\begin{center}
\includegraphics[width=1.\textwidth]{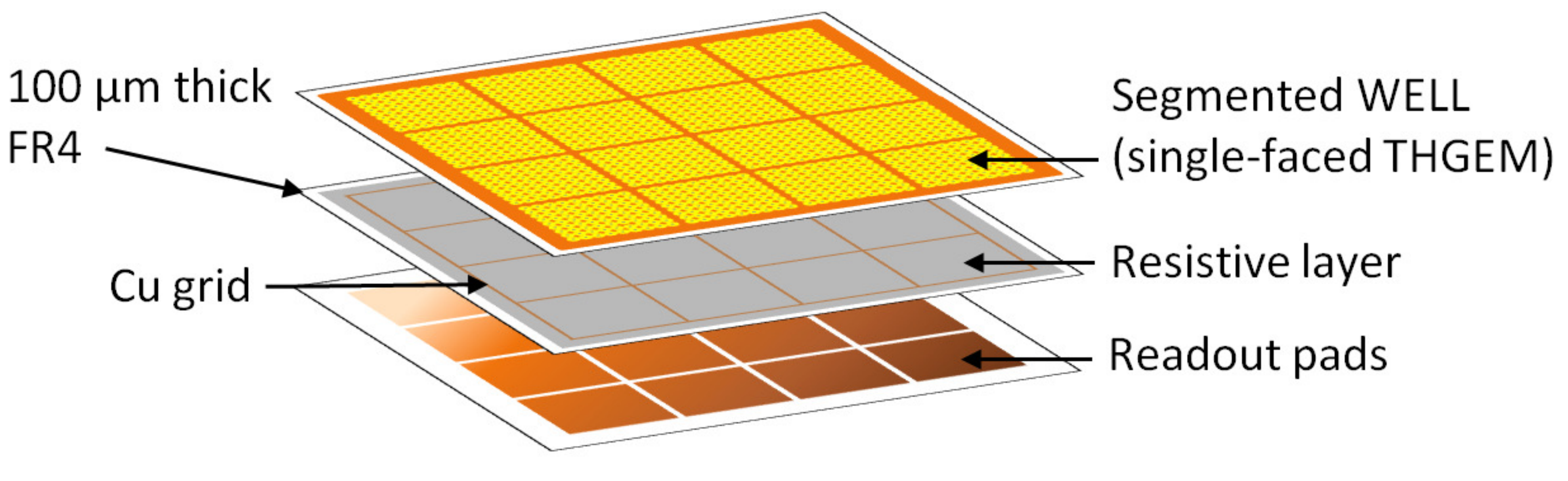}
\caption{Schematic description of Segmented Resistive Well THGEM (SRWELL) coupled to a readout pad array.}
\label{fig:SRWELL}
\end{center}
\end{figure}

A promising step forward for THGEM-based detectors can rely on configurations based on the Segmented Resistive Well THGEM (SRWELL), first suggested in~\cite{Arazi2012}. The SRWELL (figure~\ref{fig:SRWELL}) is a composite structure, comprising of a single-faced THGEM electrode (copper-clad on its top side only) coupled directly to a resistive anode in a WELL configuration ('WELL' being a THGEM with closed bottom electrode); an avalanche-induced inductive charge appears on a pad array separated from the resistive layer by a thin (100~$\rm{\mu m}$) insulating sheet. The absence of the induction gap leads to a significantly thinner geometry, while the resistive layer serves to effectively quench the energy of occasional discharges. Compared to THGEM configurations with an induction gap, higher gains are obtained for lower THGEM voltages, due to the stronger field within the closed holes~\cite{Arazi2012}. Charge spreading across the resistive layer may result in delayed inductive signals on neighboring pads and, consequently, in high pad multiplicity. This is prevented by adding a matching grid of thin copper lines below the resistive layer that allows for rapid clearance of the electrons diffusing over its surface. The SRWELL has a square-hole pattern, with 'blind' copper strips above the pad boundaries, designed to prevent discharges in holes residing directly above the metal grid lines.

We present here recent results of SRWELL-detector evaluation with 150 GeV/c muon and pion beams, conducted at the CERN SPS/H4 RD51 beam-line. Two detector configurations were studied: a 5.8-5.9 mm thick single-stage SRWELL detector and a 4.8-6.3 mm thick double-stage multiplier with a standard THGEM followed by an SRWELL. The detectors were investigated with the new Scalable Readout System (SRS) developed within CERN-RD51~\cite{Martoiu2013}. Results are presented on detection efficiency, pad multiplicity and discharge probability; future plans are discussed in brief.

\section{Experimental setup and methodology}

\subsection{The THGEM detectors}

The THGEM electrodes used in this work were $10 \times 10~\rm{cm^2}$~in size, manufactured by 0.5 mm diameter hole-drilling in either 0.4 or 0.8 mm thick FR4 plates, copper-clad on one or two sides~\cite{printe}; the holes of the double-sided THGEM electrodes were arranged in an hexagonal lattice with a pitch of 1 mm while the holes of the single-sided WELL-THGEM electrode (for the SRWELL configuration) were arranged in a square lattice with a pitch of 0.96 mm; 0.1 mm wide rims were chemically etched around the holes in both cases. The width of the 'blind' copper strips above the grid lines in the SRWELL electrodes (see figure~\ref{fig:SRWELL}) was 0.68 mm. Anodes with surface-resistivity in the range 10-20 $\rm{M\Omega}$/square were used. They were produced by spraying a mixture of graphite powder and epoxy binder on a 100 $\rm{\mu m}$~thick FR4 sheet, patterned with a square grid of 100 $\rm{\mu m}$~wide copper lines, defining an array of $10 \times 10~\rm{cm^2}$~squares (figure~\ref{fig:SRWELL}); the array corresponded to that of the $8 \times 8~\rm{cm^2}$~readout pads, patterned on a 3.2 mm thick FR4 plate. The hole-diameter, electrode thickness and rim width were chosen based on previous optimization studies~\cite{Shalem2006}. In view of previous experience with neon mixtures, offering high attainable gains at relatively low operation potentials~\cite{Azavedo2010,Cortesi2009}, the detectors were operated in Ne/5\%CH4; at normal conditions, 150 GeV muons and pions (referred to in the text as minimally ionizing particles - MIPs) deposit on the average a total number of ~60 electrons per cm along their track~\cite{Beringer2012}. 

\begin{figure}
\begin{center}
\includegraphics[width=1.\textwidth]{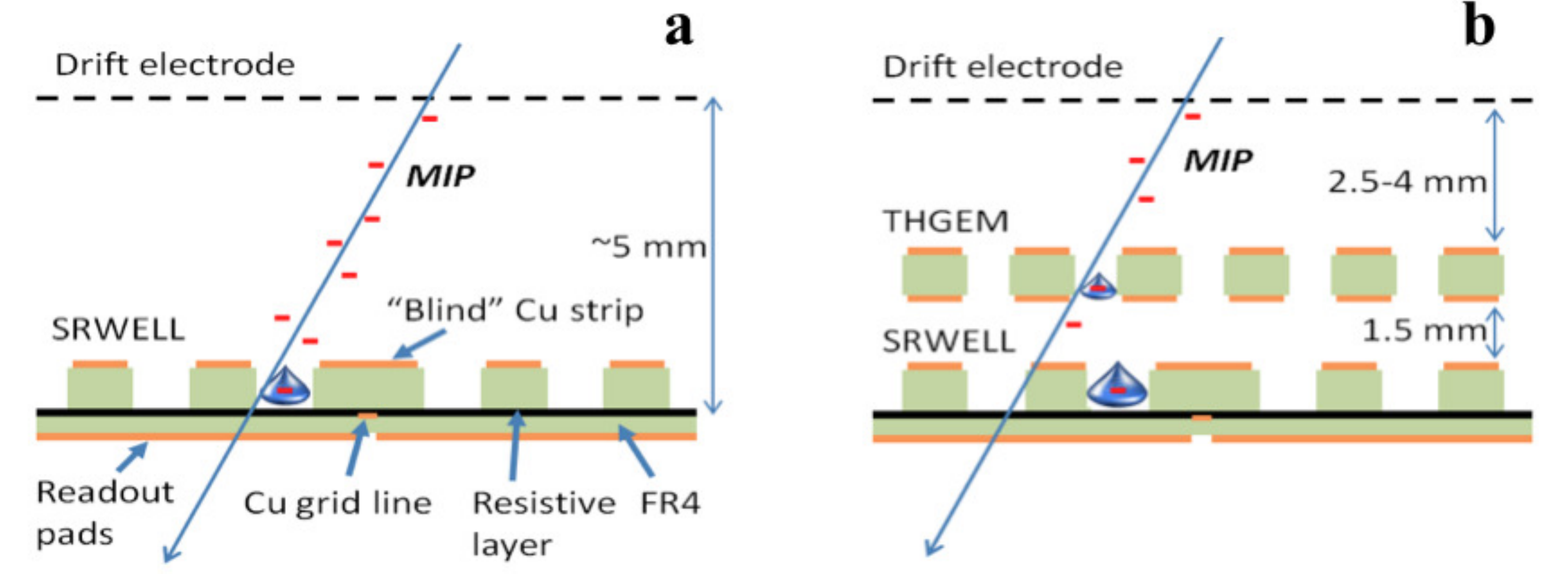}
\caption{Schematic views of the SRWELL detector (a) and the double-stage THGEM+SRWELL (b) configurations. Avalanche charges induce signals through a resistive anode, onto pads located behind a 100 $\rm{\mu m}$ thick insulator. The copper-grid below the resistive film evacuates the electrons diffusing across it surface. }
\label{fig:configurations}
\end{center}
\end{figure}

The two basic configurations investigated, with different geometrical parameters, are shown in figure~\ref{fig:configurations}. The single-stage SRWELL detector was investigated with 0.4 and 0.8 mm thick WELL electrodes; the double-stage detector, with a THGEM preceding the SRWELL, was operated with 0.4 mm thick electrodes only. Table~\ref{tab:configurations}~summarizes the geometrical parameters of both configurations and the nominal anode-resistivity values.  The drift field was maintained by setting a potential difference between the multiplier's top electrode and an additional copper-plated drift electrode (e.g. a passive THGEM) placed a few mm above.

\begin{table}
\begin{center}
\caption{Parameters of the single- and double-stage SRWELL detectors (of figure~ref{fig:configurations}) investigated in this work.}
\begin{tabular}{|c|c|c|c|c|c|}
\hline
Configuration   & Thickness  & Transfer gap  & Drift gap  & Total thickness  & Resistivity \\
			& [mm]		& [mm]		& [mm]	 &  [mm]			& $\rm{M\Omega}$/square \\
\hline
\hline
\it{Single1}       &	0.4		& -			& 	5.5	&	5.9			& 	10 \\
\it{Single2}	&	0.8		& -			&	5	&	5.8			&	10\\
\it{Double1}	&	0.4/0.4	& 1.5		&	4	&	6.3			&	20\\
\it{Double2}	&	0.4/0.4	& 1.5		&	3	&	5.3			&	20\\
\it{Double3}	&	0.4/0.4	& 1.5		&	2.5	&	4.8			&	10\\
\hline
\end{tabular}
\label{tab:configurations}
\end{center}
\end{table}

The electrodes were polarized with individual HV power-supply CAEN A1833P and A1821N boards, remotely controlled with a CAEN SY2527 unit. The voltage and current on each channel were monitored and stored. All inputs were connected through low-pass filters.

\subsection{External trigger, tracking and beam parameters }

\begin{figure}
\begin{center}
\includegraphics[width=1.\textwidth]{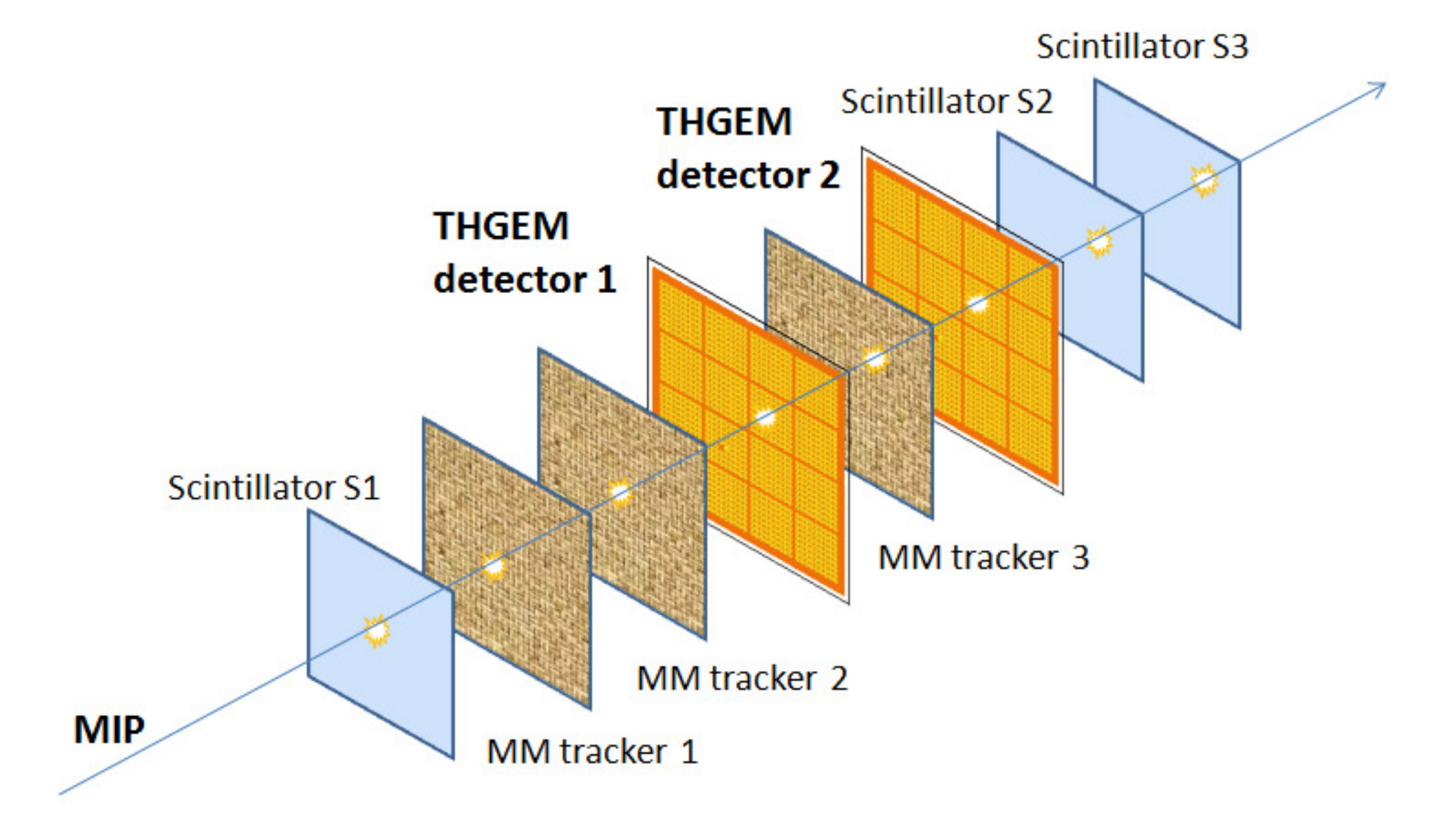}
\caption{Schematic description of the test-beam setup, incorporating the two SRWELL detectors, three scintillators and three MICROMEGAS tracking detectors. }
\label{fig:setup}
\end{center}
\end{figure}

 The external trigger system used for event selection is shown in figure~\ref{fig:setup}~\cite{Karakostas2010}; it comprised three $10 \times 10~\rm{cm^2}$~scintillators arranged in coincidence. The tracking system, covering a total area of $6 \times 6~\rm{cm^2}$, was based on position measurement with three MICROMEGAS detectors, equipped with an SRS readout system~\cite{Martoiu2013}~recording signals through APV25 front-end hybrid electronics~\cite{French2001}~similar to the one employed in our test detectors. Two chambers were tested parallel, containing different detector configurations: one was located between the two downstream trackers and the other between the third tracker and the second scintillator (figure~\ref{fig:setup}). The chambers were operated independently, but shared the same trigger and tracking system.

The experiments were performed with 150 GeV/c muons and pions. The muon beam was broad, covering the entire detector area, with a typical rate of $\sim$10-20~\Hzcm. The pion beam had a narrow central peak ($\sim$1 $\rm{cm^2}$ width) with wide, low-rate wings. Its rate was varied between $\sim$0.5  \kHzcm~and $\sim$70 \kHzcm~to investigate the detector stability and performance under different irradiation conditions.

\subsection{Readout and data acquisition}

A crucial point in the measurements was the synchronization between the data acquired with the detectors under investigation and that of the tracker. The latter was used to measure the local detection efficiency and evaluate the level of cross-talk between neighboring pads as function of the particle impact location. This was effectively achieved employing the RD51 Scalable Readout System (SRS)~\cite{Martoiu2013}.

Like the detectors of the tracker, the 64 readout pads of our detectors were coupled to APV25 chips~\cite{French2001}. The APV25, originally designed for the silicon tracking detectors in CMS, has high rate capability and low noise (equivalent of $\sim$200 electrons for a few tenth pF typical input capacitance). Due to the low noise level and high sensitivity of the chip, gas gains of the order of several 1000 were sufficient for efficient operation of our detectors.

The data acquisition of both systems was triggered by the same scintillator signals and performed with a single SRS front-end card (FEC). For each triggered event, the charge accumulated by all channels was stored. For each channel, the charge was sampled in 25 bins of 25 ns each. The mmDAQ~\footnote{Developed by M.Z.D. Byszewski (marcin.byszewski@cern.ch)}~online data acquisition software was used to store the synchronized data on a PC for further analysis.

\subsection{Analysis framework}

The track reconstruction algorithm is described in~\cite{Karakostas2010}. Tracks were selected by setting a high threshold on the MICROMEGAS, suppressing noise hits. These tracks formed the baseline objects of the analysis; the MIP trajectory through our SRWELL detectors was extrapolated from the calculated track coordinates. The detector properties, namely pulse-height distribution, efficiency and pad multiplicity (both global and local) were measured with respect to these trajectories.

Data analysis of the SRWELL detectors comprised the following two steps:
\begin{enumerate}
\item{ {\bf{Noise suppression:}} dedicated pedestal runs, without beam, were used to determine the noise level of the individual channels in each configuration. The pedestal of each channel was defined as the mean value of the charge measured in the 25 ns time intervals. The noise level of each channel was defined as the width of this distribution. Only pads with a charge signal above a threshold relative to the individual channel's noise were used in the analysis of the physics runs. A change of any of the detector parameters (HV configuration, etc.) was followed by a pedestal run.}
\item{ {\bf{Clusterization:}} neighboring pads with detected charge above the threshold formed a cluster. The cluster position was determined from the average of the pads position, weighted according to their measured charge. The charge associated with a cluster was taken as the sum of the charge of all pads in that cluster.}
\end{enumerate}

The detector efficiency was defined as the fraction of tracks where a corresponding cluster was found with its calculated position not more than 10 mm away from the track trajectory. The single-event multiplicity was defined as the number of pads in the cluster.  The average pad multiplicity was defined as the average number of pads in events where a corresponding cluster was found not more than 10 mm away from the track trajectory. 
Both the detector efficiency and pad multiplicity were measured during stable operation conditions. In particular, time intervals with detected high-voltage drops were excluded. Therefore, external parameters such as the power supply recovery time following eventual discharges did not bias the measurements. The effect of eventual discharges on the detector efficiency is discussed below.

The discharge probability was defined here as the number of discharges divided by the number of hits in the active region of the detector (i.e., in the total area covered by the crossing beam). The number of discharges was extracted directly from the power supply log files by counting the resulting drops in the monitored voltage. Since the pion beam was narrower than both the acceptance of the scintillators and the detectors, the number of pion hits in the active region of the detector was estimated as the number of triggers. For muons, where the beam was wider than the acceptance of both the scintillators and the detectors, the number of muon hits in the active area of the detector was estimated from the number of triggers, by correcting for the difference between the acceptance of the scintillators ($6 \times 6~\rm{cm^2}$) and that of the detector ($10 \times 10~\rm{cm^2}$).

\subsection{Threshold optimization}

\begin{figure}
\begin{center}
\includegraphics[width=.6\textwidth]{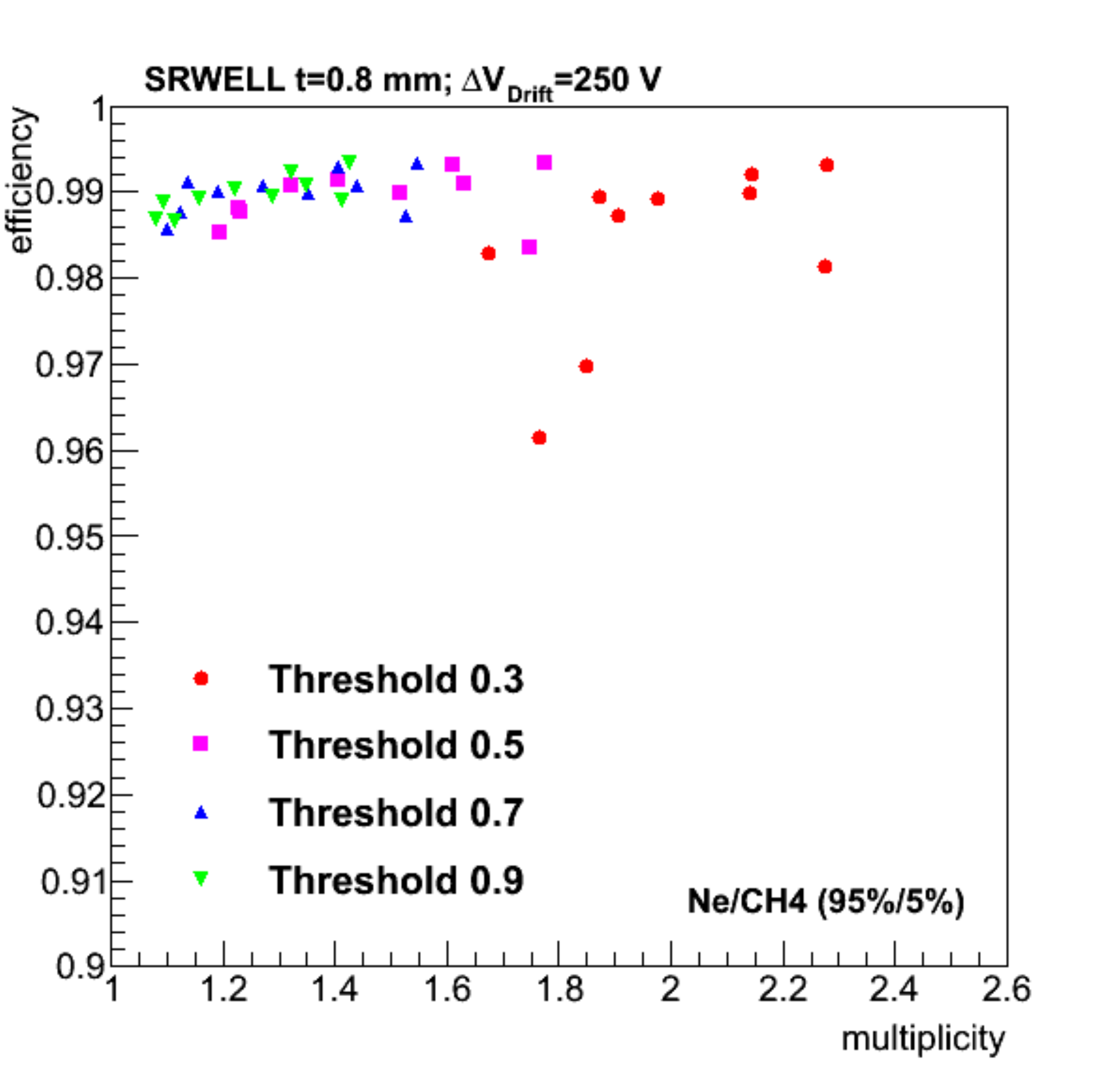}
\caption{Global detection efficiency as a function of the pad-multiplicity for different charge thresholds set in the analysis, as measured with a 0.8 thick SRWELL (figure 2a) with muons. Ne/5\%CH4; \DVDRIFT ~= 250 V.}
\label{fig:threshold}
\end{center}
\end{figure}

Unlike the tracker, where a maximal threshold was set during the analysis to ensure that only good-quality tracks are used, the threshold for the SRWELL detectors was set to optimize the detector performance.
Figure~\ref{fig:threshold}~shows an example of global detector efficiency recorded with muons, versus the pad-multiplicity; the data were recorded with a single-stage 0.8 mm thick SRWELL detector (figure~\ref{fig:configurations}a; {\it{Single2}} in table~\ref{tab:configurations}) for different thresholds set during the analysis. The detector bias voltage (\DVSRWELL) was scanned while the drift field was kept constant (\DVDRIFT ~= 250 V corresponding to a drift field of approximately 0.5 kV/cm); as explained in section 2.4, the thresholds were set relative to the noise level of the individual channels. As can be seen, relative thresholds of 0.7 or 0.9 were optimal in this case; they resulted in high detection efficiency and low pad multiplicity. A threshold of 0.7 was used throughout the analysis since it was found optimal also for the other SRWELL detector configurations investigated here.

\section{Results}

\subsection{Detection efficiency and pad multiplicity with muons}

\begin{figure}
\begin{center}
\includegraphics[width=1.\textwidth]{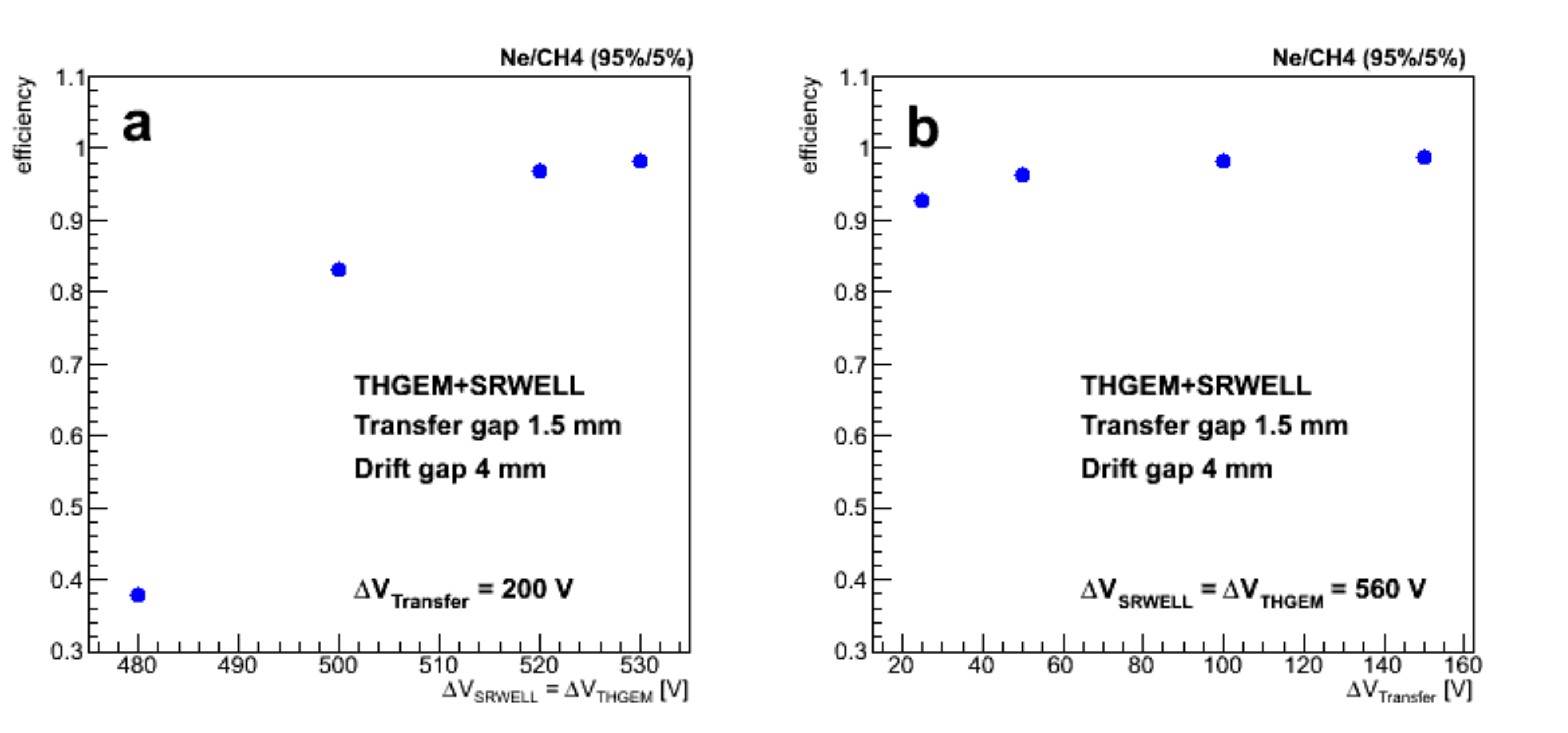}
\caption{The detection efficiency with muons of the THGEM+SRWELL in a {\it{Double1}} configuration (figure 2b; table 1) as a function of applied potential. (a) efficiency as a function of the voltage across both multipliers (\DVSRWELL ~= \DVTHGEM) at constnt \DVTRANSFER ~= 200 V; (b): efficiency as a function of the \DVTRANSFER~at constant \DVSRWELL ~= \DVTHGEM ~= 560 V.}
\label{fig:HVscanDouble}
\end{center}
\end{figure}

The detector efficiency as a function of the applied voltage is shown in figure~\ref{fig:HVscanDouble} for the THGEM+SRWELL in a {\it{Double1}} configuration (table~\ref{tab:configurations}). In figure~\ref{fig:HVscanDouble}a the same potentials were applied across the THGEM and the SRWELL (\DVTHGEM ~= \DVSRWELL). This potential was varied keeping a constant transfer potential: \DVTRANSFER ~= 200 V (corresponding to a transfer field of approximately 1.3 kV/cm); in figure~\ref{fig:HVscanDouble}b the transfer field was varied, keeping a constant potential differences across both multipliers: \DVSRWELL ~= \DVTHGEM ~= 560 V. Conditions were found for reaching close-to-unity efficiencies in a stable operation mode.

\begin{figure}
\begin{center}
\includegraphics[width=.6\textwidth]{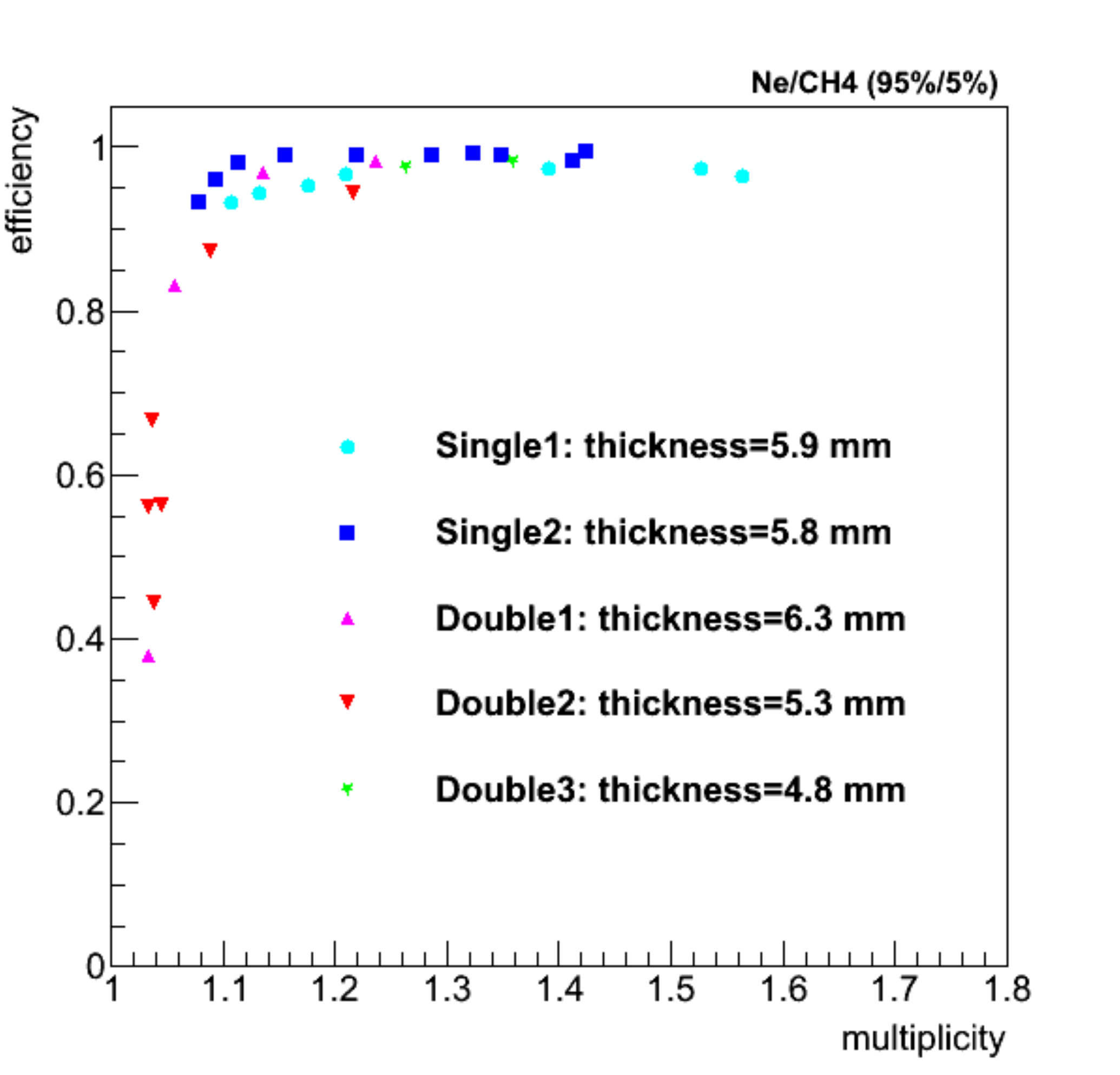}
\caption{Detection efficiency with muons as a function of the pad multiplicity measured with the single-SRWELL and the double THGEM+SRWELL configurations listed in table 1. The total detector thickness, excluding readout electronics, is provided for each case in the figure. The optimal voltages and resulting efficiency, multiplicity and gain values are given in table 2.}
\label{fig:effVsMult}
\end{center}
\end{figure}

The results of the measured detection efficiency with muons as a function of the pad multiplicity for the five configurations investigated (table~\ref{tab:configurations}) are summarized in figure~\ref{fig:effVsMult}.  All measurements were done under stable operation conditions. For both the {\it{Single1}} and {\it{Single2}} configurations, the measurements were done by increasing the SRWELL voltage; for {\it{Double1}} and {\it{Double3}} configurations the transfer voltage was kept constant (\DVTRANSFER ~= 300 V and \DVTRANSFER ~= 50 V respectively, or a transfer fields of approximately 2 kV/cm and 0.3 kV/cm) while the bias voltages of the THGEM and SRWELL were increased symmetrically (keeping \DVSRWELL ~= \DVTHGEM ); for {\it{Double2}} the transfer voltage was kept constant (\DVTRANSFER ~= 225 V, or approximately transfer field of 1.5 kV/cm) while the bias voltages of the THGEM and SRWELL were scanned asymmetrically. High detection efficiencies (>95\%) and low pad multiplicity values (<1.3) were recorded with all configurations. Note that over 98\% efficiency at multiplicity values below 1.2 were measured with the {\it{Single2}} and {\it{Double1}} detector configurations. The applied voltages and resulting gains at optimal detection conditions (lowest pad multiplicity at the efficiency plateau) are summarized for all configurations in table~\ref{tab:results}.

\begin{table}
\begin{center}
\caption{The applied potentials and resulting effective gain at the optimal detection conditions (lowest pad multiplicity at detection efficiency plateau) of all of the tested configurations.}
\begin{tabular}{|c|c|c|c|c|c|c|}
\hline
Configuration   & \DVSRWELL  & \DVTRANSFER  & \DVTHGEM  & Efficiency  & Multiplicity & Effective \\
			& [V]			& [V]				& [V]	 	& 	 [mm]   &  		& gain \\
\hline
\hline
\it{Single1}       &	610		&	-		& - 			& 	0.97		&	1.2	&	1200 \\
\it{Single2}	&	780		&	-		& - 			& 	0.98		&	1.1	&	2000 \\
\it{Double1}	&	530		&	300		& 530 		& 	0.98		&	1.2	&	6500 \\
\it{Double2}	&	540		&	225		& 540 		& 	0.95		&	1.2	&	8200 \\
\it{Double3}	&	540		&	50		& 540 		& 	0.98		&	1.3	&	4000 \\
\hline
\end{tabular}
\label{tab:results}
\end{center}
\end{table}

\subsection{Local characteristics (muons)}

\begin{figure}
\begin{center}
\includegraphics[width=1.\textwidth]{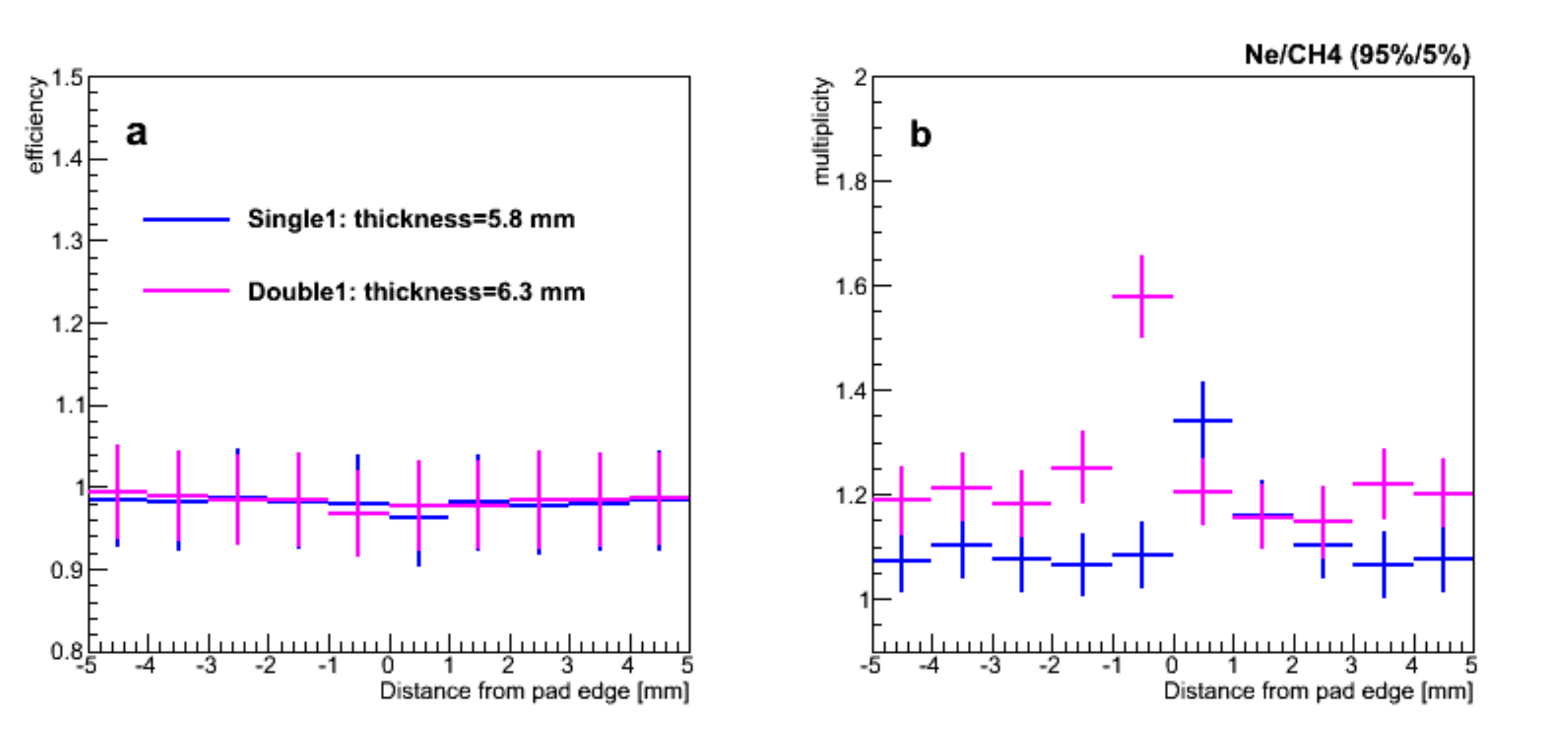}
\caption{Local efficiency and multiplicity data measured with the Single2 SRWELL and the Double1 THGEM+SRWELL, using the tracking system. (a) Detection efficiency as a function of distance from the edge of the pad.  (b) Pad multiplicity as a function of distance from the edge of the pad. The operation potentials are depicted in table 2.}
\label{fig:local}
\end{center}
\end{figure}

The accurate tracking allowed for studying the local characteristics of the detector. For each configuration, an optimal working point (lowest multiplicity at the efficiency plateau) was selected. At these conditions, the local efficiency and multiplicity were measured as a function of the particle impact distance from the edge of the pad. The local efficiency and multiplicity plots of the {\it{Single2}} and {\it{Double1}} detector configurations are shown in figure~\ref{fig:local}. As expected, close to the edge of the pad the charge is shared between two neighboring pads. Similar behavior was measured with the other three configurations.

As discussed in our previous work~\cite{Arazi2012} the charge sharing also depends on the detector configuration, the anode type and its resistivity; in the SRWELL, the metal strips below the resistive layer prevent charge propagation to neighboring pads and the charge sharing is due mostly to the event induced avalanche size. By definition, the charge sharing between neighboring pads results in higher average pad multiplicity. This suggests that the pad multiplicity provides indeed additional information concerning the track position, which could be exploited.  Since in such events, less charge is induced on each pad, the signal-to-noise separation is worse, resulting in slightly lower detection efficiency.

\subsection{Operation with pions}

\begin{figure}
\begin{center}
\includegraphics[width=1.\textwidth]{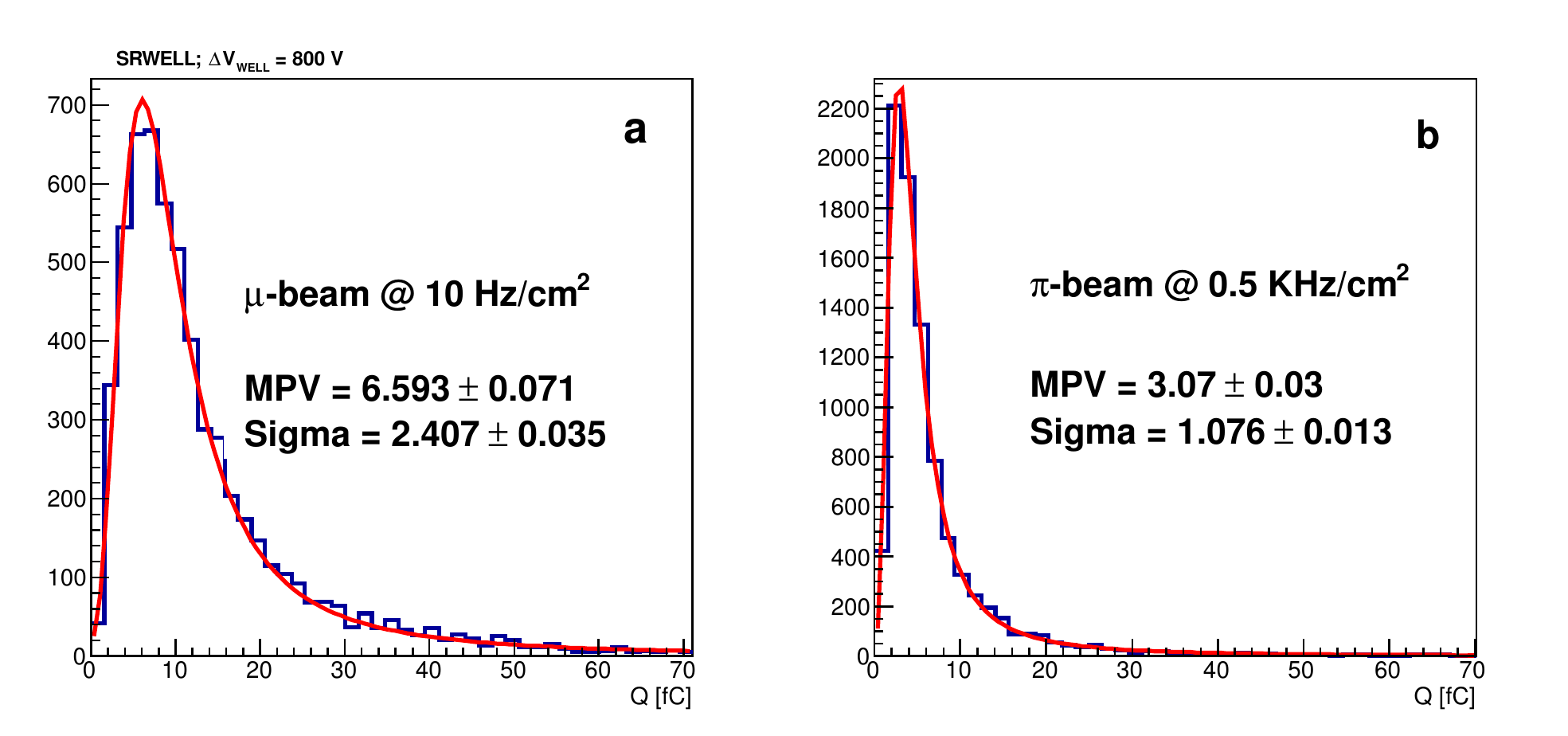}
\caption{Cluster charge distributions measured with the {\it{Single2}} configuration (table 1) at an operation voltage \DVSRWELL ~= 800 V (a) muons, 10-20 \Hzcm;  (b) pions, 0.5 \kHzcm. The measured data (blue histograms) are fitted to Landau distributions.}
\label{fig:muPiSingle}
\end{center}
\end{figure}

\begin{figure}
\begin{center}
\includegraphics[width=1.\textwidth]{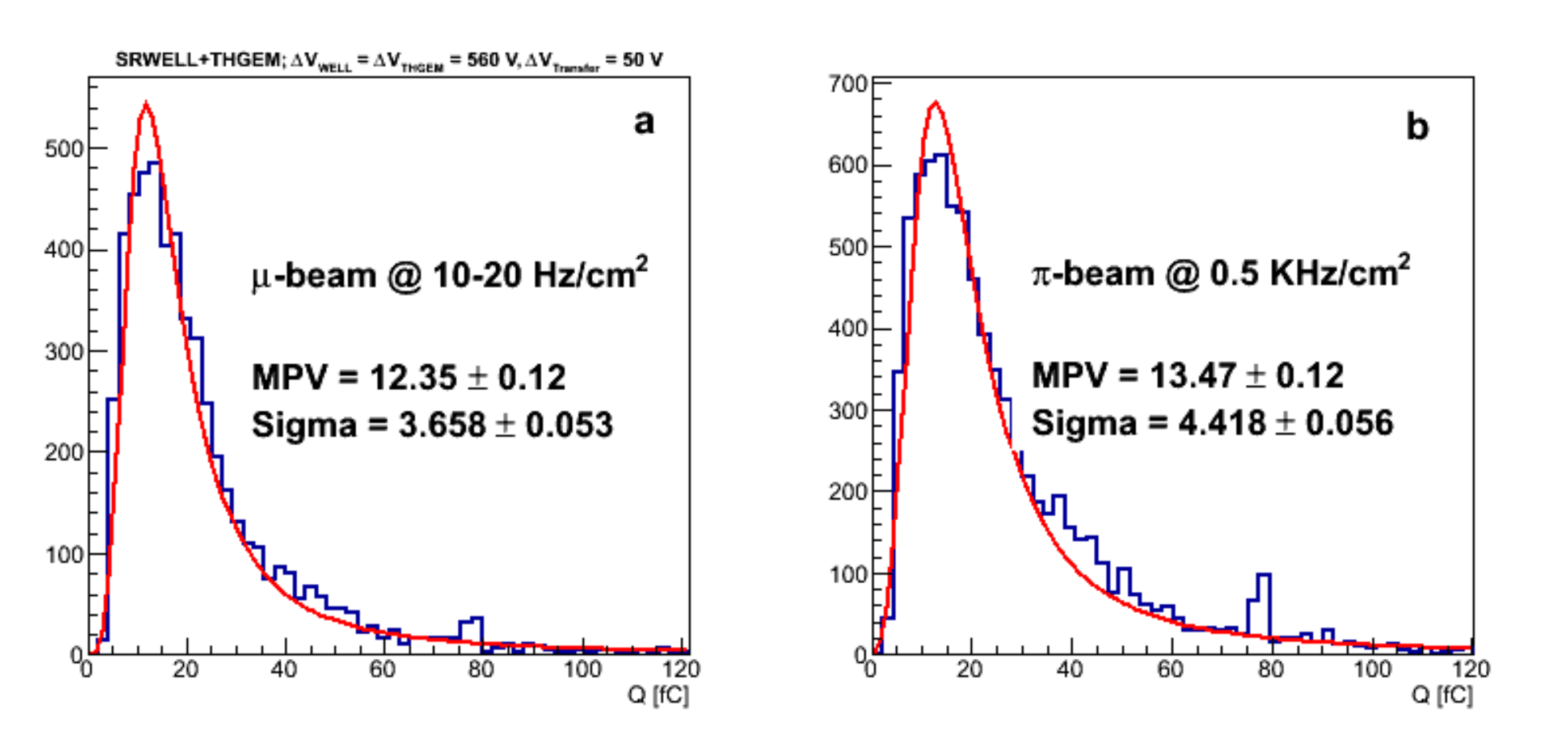}
\caption{Cluster charge distributions measured with the {\it{Double1}} configuration (table 1) at operation voltages \DVSRWELL ~= 560 V, \DVTHGEM ~= 560 V and \DVTRANSFER ~= 50 V. (a) muons, 10-20 \Hzcm;  (b) pions, 0.5 \kHzcm. The measured data (blue histograms) are fitted to Landau distributions.}
\label{fig:muPiDouble}
\end{center}
\end{figure}

Experiments were carried out with 150 GeV/c pions, to investigate the detectors' behavior under higher event rate and in the presence of potential higher-ionization hadron-induced background. Examples of cluster charge distributions (defined in section 2.4) measured with both muons and pions in the {\it{Single2}} and {\it{Double1}} configurations are shown in figure~\ref{fig:muPiSingle} and figure~\ref{fig:muPiDouble} respectively. For each configuration the measurements were carried out with the same high-voltage configuration for both beam types; \DVSRWELL ~= 800 V for {\it{Single2}}. \DVSRWELL ~= \DVTHGEM ~= 560 V and \DVTRANSFER ~= 50 V for {\it{Double1}}. The measured data (blue histograms) are fitted to Landau distributions (red curves). As can be seen, the measured cluster charge distribution of the double-stage configuration (figure~\ref{fig:configurations}b) remained practically unchanged in the transition from muons to pions; however, the operation of the single-stage configuration (figure~\ref{fig:configurations}a) with pions resulted in an average-gain drop of over 50\%, as indicated by the drop of the Landau Most Probable Value (MPV) from 6.6 fC to 3.1 fC.

The substantial drop of the MPV in the single-SRWELL irradiated with pions (figure~\ref{fig:muPiSingle}b) is yet unclear, requiring further investigations. The possibility of rate-dependence of the gain is ruled out, since this gain drop is not reproduced in the lab when irradiating the detector with an x-ray source at the same rates. Note that some high-voltage fluctuations that appeared at the much higher rates in the single-stage configuration operated with pions could have also caused this effect. Careful studies are in course.

\section{Discharge analysis}

In previous laboratory studies, occasional-discharge rates and amplitudes measured with a resistive (but not segmented) WELL were considerably lower compared to that of THGEM and WELL configurations coupled to metal pads~\cite{Arazi2012}. In the present beam study, two beam-related types of discharges were observed. Large voltage drops of  50-150 V at either one or more of the detector electrodes, characterized the first type; a few volts drop characterized the second type, referred in the text as micro-discharges.

The probability of large discharges was low, < $10^{-6}$, for all of the investigated configurations. They depended on the detector configuration, on the applied voltage and on the particle flux. In particular, no large discharges of the SRWELL electrode were observed with the double-stage configurations. Large discharges were followed by long recovery times of the power supply and possibly of the detector. In some cases, the SRS readout electronics had to be reconfigured or power-cycled.

\subsection{The effect of micro-discharges}

\begin{figure}
\begin{center}
\includegraphics[width=.6\textwidth]{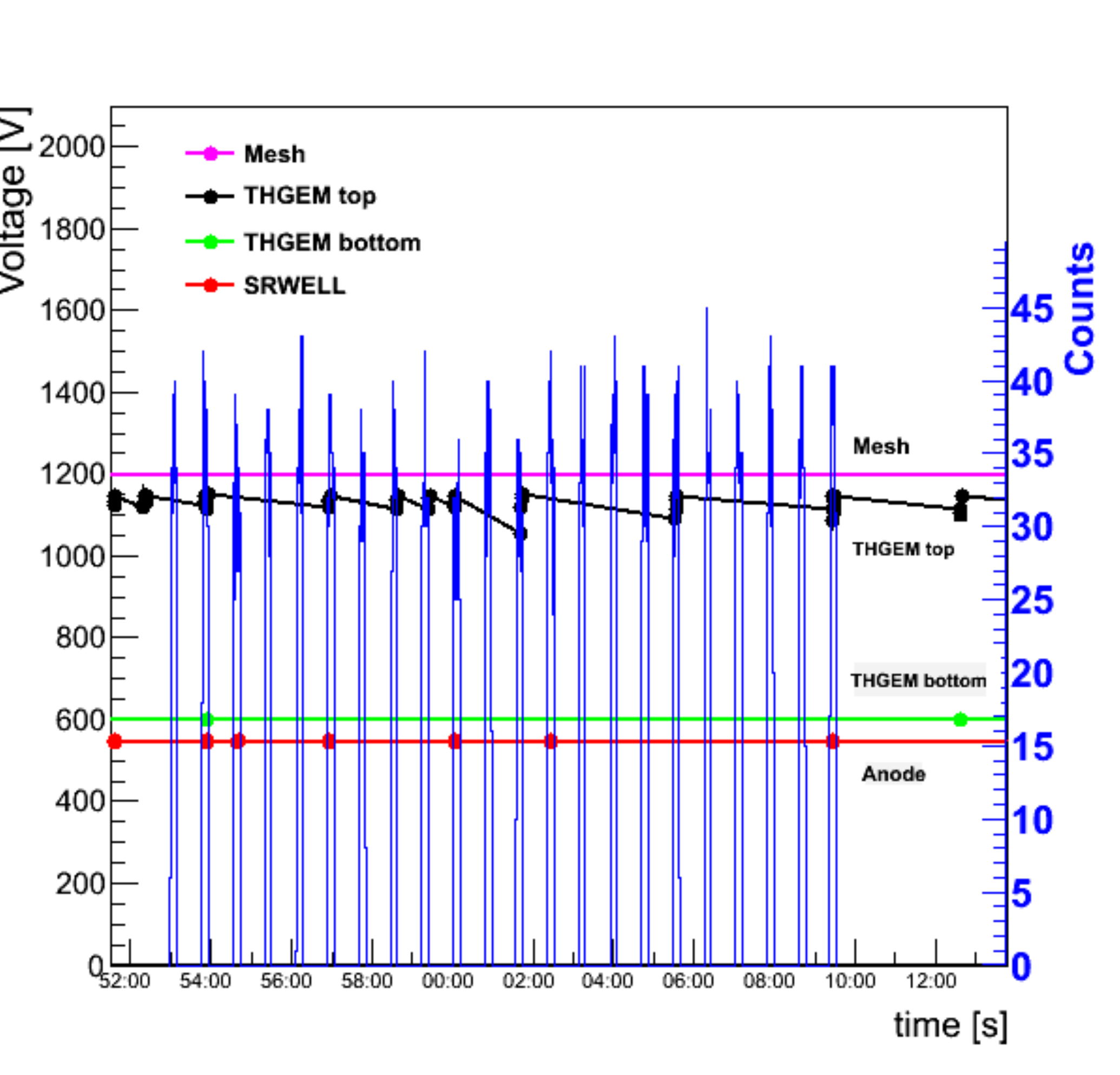}
\caption{The {\it{Double3}} THGEM+SRWELL detector configuration. High-voltage variation in time overlaid on the spill structure, as measured in a pion beam at a rate of 3 \kHzcm. No high voltage variations were measured on the mesh (magenta curve). The small variations (shown as dots) measured on the THGEM top electrode, THGEM bottom electrode and WELL electrode are shown in black, green and red curves respectively. }
\label{fig:beamHVprofile}
\end{center}
\end{figure}

Typical high-voltage variations in time, measured on the different electrodes of a double-stage configuration ({\it{Double3}}) operated in a pion beam, are shown in figure~\ref{fig:beamHVprofile}. In order to study the correlation between the occasional voltage drops and the beam flux, the variations are overlaid on the number of tracks measured every ten seconds (the beam consisted of 10 seconds spills every 50 seconds). As can be seen, the THGEM top and SRWELL electrodes are those with the highest activity. The THGEM top displays 50-100 V discharges, while the SRWELL shows only micro-discharges ($\sim$3 V drops). Discharges are mainly, but not exclusively, correlated with the beam. Discharges on the THGEM top and micro-discharges on the SRWELL are partially correlated.

\begin{figure}
\begin{center}
\includegraphics[width=.6\textwidth]{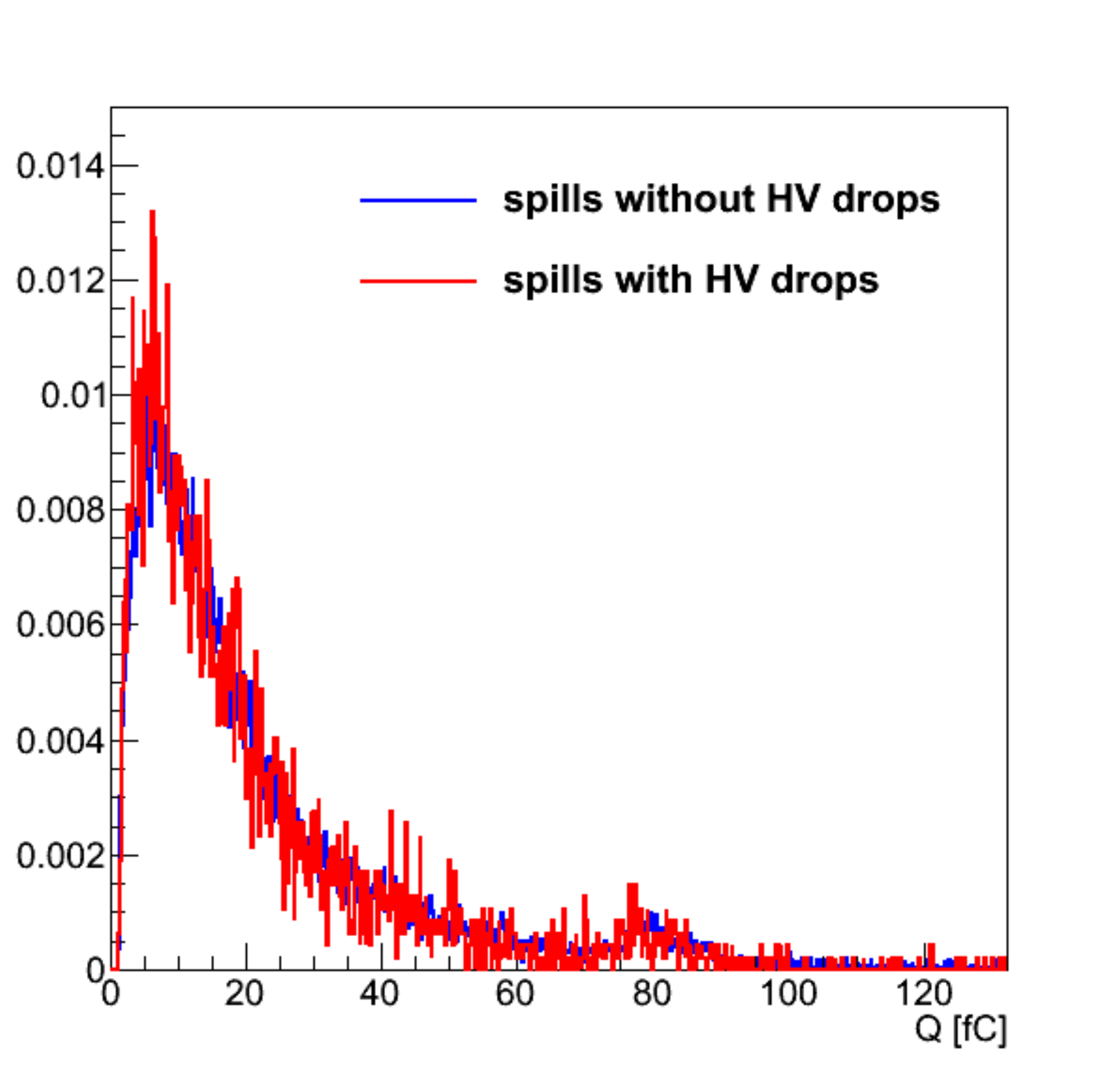}
\caption{Cluster charge distributions recorded in the {\it{Double3}} THGEM+SRWELL detector (of figure 10) with pions at a rate of 3 \kHzcm.  The distributions were measured during spills with (red) and without (blue) micro-discharges in the same operation conditions: \DVSRWELL ~= 550 V, \DVTHGEM ~= 550 V and \DVTRANSFER  ~= 50 V resulting in an effective gain of $\sim$2500.}
\label{fig:clusterMicroDischarge}
\end{center}
\end{figure}

The performance of the double-stage configuration ({\it{Double3}}) was measured during pion-beam spills, with and without the appearance of micro-discharges. Similar (within statistical fluctuation) efficiencies were measured in both cases. The very similar cluster-charge distributions recorded in both conditions are shown in figure~\ref{fig:clusterMicroDischarge}. We conclude that the effect of micro-discharges on the performance of the double-stage detector is negligible. Similar analysis could not be made with the single-stage detectors since it was difficult to find spills with and without high-voltage drops at the same high-voltage configuration.

\section{Summary and Discussion}
        
This work focused on the test-beam results with our thinnest (<6.3 mm) THGEM/SRWELL detectors, as possible candidates for Digital Hadron Calorimetery applications. A total of five configurations were investigated; two single-stage SRWELL and three double-stage THGEM + SRWELL configurations. All the detectors were operated in stable conditions at typical effective gains in the range 1000-8000; signals were processed with frontend APV25 chips and SRS electronics. Detailed results of systematic studies of processes in the different detector configurations discussed here will be the subject of forthcoming publications.
	
High detection efficiencies (over 95\%) and low average pad-multiplicity values (less than 1.2) were recorded with all the configurations investigated here. In particular, efficiencies above 98\% with an average pad-multiplicity below 1.2 were recorded with both the {\it{Single2}} and {\it{Double1}} configurations, with a total thickness of 5.8 and 6.3 mm respectively. In the transition from low-rate muon-beam to high-rate pion beam a significant, yet unclear, gain drop was observed with the single-stage configuration, at the same operation voltage; the effect is being investigated. No gain drop was observed with the double-stage configurations.

Two types of occasional discharges were observed; large discharges, characterized by a voltage drops of 50-150 V, were observed with the single-stage configurations at low appearance rates (< $10^{-6}$), and very rarely on the top THGEM electrode of the double-stage configurations. Micro-discharges, characterized by a voltage drop smaller than 10 V, were recorded in all configurations; these did not affect the cluster-charge distributions and the detection efficiency of the double-stage configurations. Further studies are required in order to understand the effect of micro-discharges on the single-stage configuration.
	
Compared to the technologies already explored as potential sampling-elements for future Digital Hadron Calorimeters, the performance of the $10 \times 10 ~\rm{cm^2}$~THGEM-based detectors reported in this work is better than that of $1 \times 1~ \rm{m^2}$~RPCs and $30 \times 30~\rm{cm^2}$~GEM detectors and similar to that of $1 \times 1~\rm{m^2}$~MICROMEGAS. The detector thickness can be further reduced by optimizing the different gaps. The thinner single-stage configuration has additional advantages in term of cost and complexity of the detector. 

The proportional response of our THGEM-based sampling elements makes them attractive also for the semi-DHCAL concept. Note that preliminary investigations of a single-THGEM and SRWELL yielded satisfactory results, in terms of efficiency and pad multiplicity \footnote{A collaborative ongoing effort with LAPP annecy}, employing the MICROROC readout electronics; the latter (developed for DHCAL applications ~\cite{Adloff2012}) can be operated also in dual-threshold semi-digital mode.

The challenging task of developing a large-scale thin single-stage detector also requires resolving the observed gain-drop effect. It will be the focus of our future work. The assembly of larger $30 \times 30~\rm{cm^2}$~SRWELL-detector prototypes is in course.

\acknowledgments

 We are indebted to Hans Muller, Sorin Martoiu, Marcin Byszewski and Konstantinos Karakostas for their assistance with the SRS electronics and associated software. We thank Victor Revivo and Amnon Cohen of Print Electronics Israel for producing the detector electrodes. We thank Rui de Oliveira of CERN for helpful discussions and assistance with the detector electrodes, for producing the readout pad array and for his on-site support during the beam test. We thank Andy White of UTA for his interest in this work and his support and Leszek Ropelewski and his team at CERN's Gas Detector Development group, in particular George Glonti, for their kind assistance during the tests. This work was supported in part by the Israel-USA Binational Science Foundation (Grant 2008246), by the Benozyio Foundation and by the FCT Projects PTDC/FIS/113005/2009 and CERN/FP/116394/2010. The research was done within the CERN RD51 collaboration. H. Natal da Luz is supported by FCT grant SFRH/BPD/66737/2009. C. D. R Azevedo is supported by the SFRH/BPD/79163/2011 grant. A. Breskin is the W.P. Reuther Professor of Research in the Peaceful use of Atomic Energy.

\end{document}